\documentclass[preprint]{ptephy}
\preprintnumber{XXXX-XXXX} 
\usepackage{graphicx} 
\usepackage{color}

\begin{document}

\title{Development of a low-mass and high-efficiency charged particle detector}

\author{\name{D.~Naito}{1\ast},
\name{Y.~Maeda}{1}\thanks{Kobayashi-Maskawa Institute, Nagoya University, Nagoya 464-8602, Japan.},
\name{N.~Kawasaki}{1},
\name{T.~Masuda}{1}\thanks{Present address : Research Core for Extreme Quantum World, Okayama University, Okayama 700-8530, Japan.},
\name{H.~Nanjo}{1},
\name{T.~Nomura}{2},
\name{M.~Sasaki}{3}
\name{N.~Sasao}{4},
\name{S.~Seki}{1},
\name{K.~Shiomi}{1}\thanks{Present address : Institute of Particle and Nuclear Studies, High Energy Accelerator Research Organization (KEK), Tsukuba, Ibaraki 305-0801, Japan.},
\name{Y.~Tajima}{3}
}

\address{\affil{1}{Department of Physics, Kyoto University, Kyoto 606-8502, Japan}
\affil{2}{High Energy Accelerator Research Organization (KEK), Ibaraki 305-0801, Japan}
\affil{3}{Department of Physics, Yamagata University, Yamagata 990-8560, Japan}
\affil{4}{Research Core for Extreme Quantum World, Okayama University, Okayama 700-8530, Japan}
\email{d.naito@scphys.kyoto-u.ac.jp}}

\begin{abstract}%
We developed a low-mass and high-efficiency charged particle detector for an experimental study of the rare decay $K_L \rightarrow \pi^0 \nu \bar{\nu}$.
The detector is important to suppress the background with charged particles
to the level below the signal branching ratio predicted by the Standard Model (O(10$^{-11}$)).
The detector consists of two layers of 3-mm-thick plastic scintillators with wavelength shifting fibers embedded and Multi Pixel Photon Counters for readout.
We manufactured the counter and evaluated the performance such as light yield, timing resolution, and efficiency.
With this design, we achieved the inefficiency per layer against penetrating charged particles to be less than $1.5 \times 10^{-5}$,
which satisfies the requirement of the KOTO experiment determined from simulation studies.

\end{abstract}

\subjectindex{xxxx, xxxx}

\maketitle
 
\section{Introduction}
$K_L \rightarrow \pi^0 \nu \bar{\nu}$ is a rare decay that directly violates CP symmetry.
Its branching ratio is predicted with a small theoretical uncertainty to be $2.43(39)(6) \times 10^{-11}$ in the Standard Model (SM)  of elementary particle physics~\cite{PhysRevD.83.034030},
where the numbers in the first and second parenthesis indicate parametric and intrinsic uncertainties, respectively.
Several new physics models suggest large enhancement up to a few orders of magnitude in the branching ratio,
and $K_L \rightarrow \pi^0 \nu \bar{\nu}$ is a powerful tool for searching for new physics~\cite{BSM}.
Experimentally, this decay has not yet been observed and its upper limit of $2.6\times 10^{-8}$ (90\% CL) was given so far (KEK-E391a~\cite{E391A}).

The KOTO experiment~\cite{KOTO} is an experiment dedicated to observing the $K_L \rightarrow \pi^0 \nu \bar{\nu}$ decay for the first time,
using an intense kaon beam at Japan Proton Accelerator Research Complex (J-PARC)~\cite{JPARC}.
We use an upgraded detector from the KEK-E391a which was the pilot experiment for the KOTO experiment.
The KOTO experiment was designed to achieve the single event sensitivity of $8\times10^{-12}$. 

\begin{figure*}[hbt]
\begin{center}
	 \includegraphics[width=0.8\textwidth]{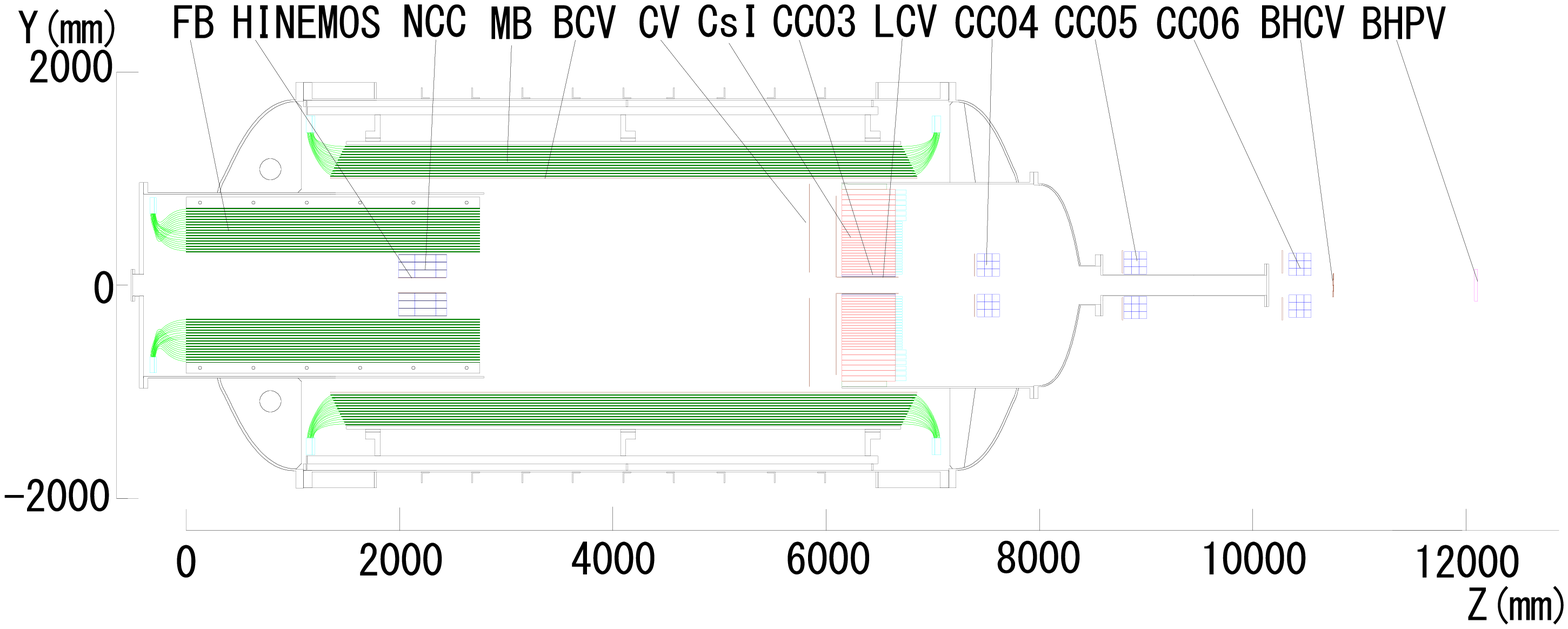}
	 \caption
	 {Cross sectional view of the KOTO detector. 
	  CV in the figure indicates the charged particle counter we developed.
	  Most of detectors, including the CV, are installed in a cylindrical vacuum chamber.
	 We search for the $K_L \rightarrow \pi^0 \nu \bar{\nu}$ decay in flight that occurs in the central region of the vacuum chamber.}
	\label{fig:KotoDetector}
\end{center}
\end{figure*}

Figure~\ref{fig:KotoDetector} shows a schematic view of the KOTO detector. 
The signature of the $K_L \rightarrow \pi^0 \nu \bar{\nu}$ consists of two photons from a $\pi^0$ decay with no other particles detected. 
The position and energy of each of the two photons are measured by a CsI electromagnetic calorimeter.
The decay-vertex position and the transverse momentum of $\pi^0$ are reconstructed to identify the decay.
Hermetic neutral and charged particle detectors (veto detectors) surround the decay region in order to ensure that no extra particles exist.
The hermetic detector system plays an essential role to achieve a background rejection factor of $10^{11}$ required for the experiment.
The Charged Veto (CV) is located in front of the CsI electromagnetic calorimeter, and detects charged particles entering the calorimeter.
Almost all the charged decay modes of $K_L$ have two charged particles; if two charged particles fly toward the CsI calorimeter and are misidentified as two gammas, such an event can be a background
\footnote{For the $K_L \rightarrow \pi^+ \pi^- \pi^0$ decay,
if $\pi^-$ and $\pi^+$ have a charge exchange interaction ($\pi^-p\rightarrow \pi^0n$ or $\pi^+n\rightarrow \pi^0p$) before $\pi^-$ or $\pi^+$ are detected by the CV,
and particles at final states are not detected by any veto counters, $K_L \rightarrow \pi^+\pi^-\pi^0$ can be backgrounds.}.
There are about $3 \times 10^{10}$ times more $K_L$ decays that include charged particles than the expected signal.
The CV is required to reject such background events by more than a factor of $10^{10}$, after other detectors are used to suppress the background.
To achieve this goal, we installed two layers of plastic scintillators as the CV to cover the upstream surface of the calorimeter. 
The two charged particles from the $K_L$ mainly make four hits on the two layers of the CV.
Because the total reduction is expected to be  the fourth power of the inefficiency of each layer,
we can achieve the required background reduction
if the charged particle inefficiency is less than 10$^{-3}$ for each layer.
We developed and evaluated a new detector for the CV to achieve the requirement.

\section{CV design and production}
\subsection{Concepts}
We used plastic scintillator strips with wavelength-shifting (WLS) fibers  embedded for the CV. 
Strips of these scintillators are arranged to cover the whole region of the calorimeter.
The segmentation is necessary to reduce the counting rate per readout channel
\footnote{From the simulation, the maximum hit rate of the scintillator strips is less than 250 kHz at the designed beam power of the KOTO experiment.}.
 It was designed with three functionalities: minimal interaction, small inefficiency, and good timing resolution.

First, the low mass feature is required to minimize neutron interaction at the CV because of the following reason.
At the neutral beam line for the KOTO experiment (KL beam line), a small fraction of neutrons remains in the halo region of the beam.
If a neutron interacts with the CV and generates a $\pi^{0}$ or $\eta$,
two photons from the $\pi^{0}$ or $\eta$ decay can fake a signal.
To reduce such events, the mass of the CV must be as low as possible.

Second, small inefficiency is essential to achieve the KOTO goal.
In other words, high light yield should be maintained, and there should be no gaps between the scintillator strips.
The CV needs to have more than 10~photoelectrons (p.e.) per $100~\mathrm{keV}$
energy deposit to achieve an inefficiency level of less than 10$^{-3}$ at a 100 keV threshold
\footnote{The $100~\mathrm{keV}$ threshold enables the CV to detect charged pions which pass through the scintillator by more than 0.5 mm before a charge exchange interaction,
with the inefficiency less than $10^{-4}$\cite{KOTO}. 
A counting loss occurs when a charged pion makes the charge exchange interaction ($\pi^+ n \rightarrow\pi^0p$ or $\pi^- p \rightarrow\pi^0n$) and gives a small energy deposit in the scintillator.
This threshold is small enough to suppress the background from the $K_L \rightarrow \pi^+ \pi^- \pi^0$ decay.} 
, according to our simulation study.
If we reduce the thickness of the scintillator to suppress the interaction of the neutrons, the light yield decreases because the number of reflections increases and the light is attenuated at every reflection.
This would increase the inefficiency.
Using WLS fibers embedded in the plastic scintillators enabled the CV to reduce the effects of attenuation.
The fibers absorb blue light ($\sim430~\mathrm{nm}$) from the scintillator and emit green light ($\sim480~\mathrm{nm}$), which is expected to have a longer attenuation length.
The light from the fibers are read at both ends by Multi Pixel Photon Counters (MPPCs).
MPPCs have an approximately 2.5 times higher detection efficiency for green light ($\sim45\%$ at $480~\mathrm{nm}$~\cite{MPPC}) than ordinary photomultipliers.
The arrangement of the scintillator strips without gaps is important to minimize
the inefficiency, which will be discussed later.

Third, the CV is required to have a good timing resolution.
A  poor timing resolution requires a wider timing window to keep a sufficient efficiency, 
and causes event loss due to accidental hits.
The timing resolution is required to be less than $3~\mathrm{ns}$ to suppress the event loss to lower than 5\%.

\subsection{CV structure}\label{sec:CVstructure}
\begin{figure}[!htb]
\begin{center}
	 \includegraphics[width=12cm,clip]{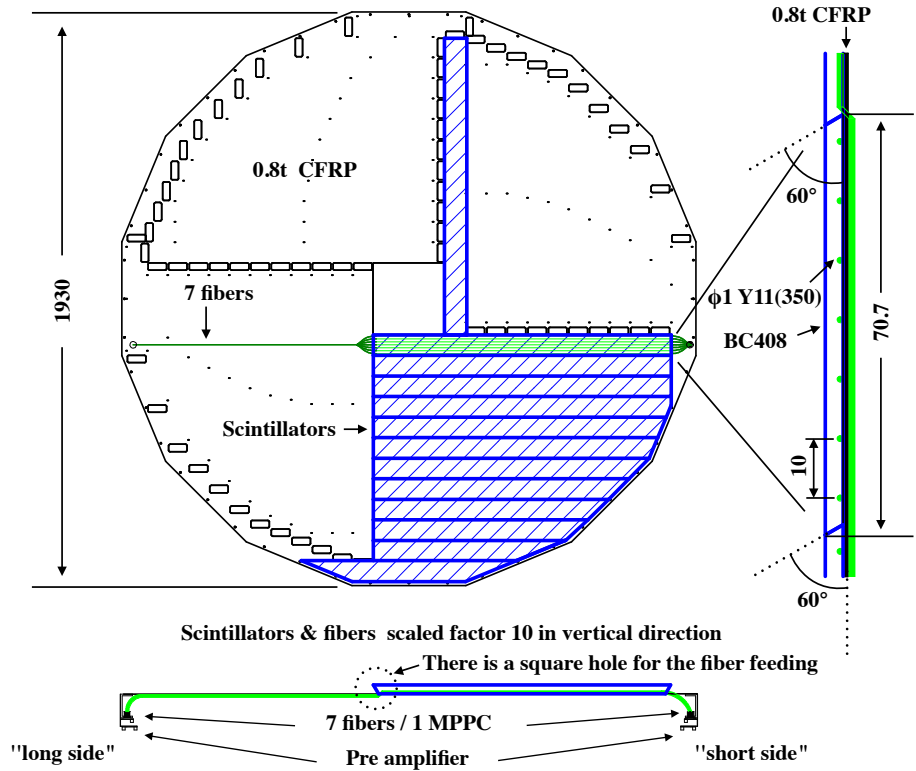}
		\caption{Schematic view of a CV front layer.
		The figure at the bottom shows the side view of the scintillator near the center viewed from the bottom.
		We call each end of a strip ``long side'' and ``short side'' according to the length of fibers extending out of the strip.
		In the bottom figure, ``long side'' is the left side and ``short side'' is the right side of the strip.
		The figure on the right shows the side view of the scintillator viewed from the right.
		Each layer has a quadrant structure with a 90-degree symmetry.
		Plastic scintillator strips are tied onto an octagonal supporting CFRP plate with fluorocarbon wires passing through the cutouts in the edges of strips.
		Rectangular holes on the CFRP plate are for routing WLS fibers to MPPCs at the outer circumference. 
                 } 
		\label{fig:CVDesign}
	\end{center}
\end{figure}

The CV consists of two layers of plastic scintillators. Each layer is named as ``front'' and ``rear'' along the beam direction.
Figure~\ref{fig:CVDesign} shows a schematic view of the CV layer. 
\begin{figure}[!htbp]
	\begin{center}
	 \includegraphics[width=8cm,clip]{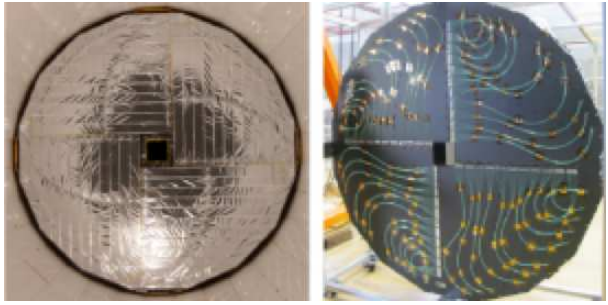}
	 \caption{Photographs of the CV layers. 
	 The left photograph shows an upstream view of the front layer. The scintillator strips are wrapped with aluminized mylar sheets. 
	 The right photograph shows a downstream view of the rear layer.
	 The wavelength-shifting fibers, which come from the scintillators on the upstream side of the CFRP plate through the holes, are routed to MPPCs on the outer supporting ring.
	 } 
	\label{fig:CVPhoto}
\end{center}
\end{figure}

Each layer has a quadrant structure with a 90-degree symmetrical arrangement, with a beam hole in the center. 
Each quadrant consists of twelve (eleven) scintillator strips in the front (rear) layer.
Each scintillator is 70.7~mm wide and 3~mm thick.
The lengths of the strips are determined so as to make the outer shape an octagon, as summarized in Table~\ref{tab:CVConfig}.
The strips in a layer are tied onto a 0.8-mm-thick carbon-fiber-reinforced-plastic (CFRP) plate with fluorocarbon wires passing through the cutouts in the edges of the strips.
The cutouts of the neighboring scintillator strips overlap and make a $0.6~\mathrm{mm} \times 0.6~\mathrm{mm}$ insensitive region.
The ratio of the inefficient area to the whole area is 10$^{-5}$ for all the cutouts, and is smaller than the inefficiency requirement ($<10^{-3}$).
In order to minimize gaps between strips, the four sides of each strip are cut at an angle of 60~degrees as shown in Fig.\ref{fig:CVDesign}, 
so that neighboring strips can be arranged with an overlap without increasing thickness.
\begin{table}[!h]
\begin{minipage}{0.5\hsize}
	 \begin{center}
	 \caption{Summary of the CV structure.}
	\scalebox{0.75}{
	\begin{tabular}{lrr} \hline	
		  & Front layer & Rear layer \\ \hline 	
		Outer radius (mm) & 950 &  850\\
		Beam hole size (mm $\times$ mm) &  $ 242 \times 242$ & $154 \times 154$\\
		Number of scintillator strips & 48 ($12 \times 4$) & 44 ($11 \times 4$) \\ 
		Length of the strips (mm) & 644--1002 & 490--917\\ 
		Number of readout channels & 96 & 88 \\ \hline
	\end{tabular}}
	\label{tab:CVConfig}
	\end{center}
\end{minipage}
\begin{minipage}{0.5\hsize}
	 \begin{center}
	  \caption{Summary of the MPPC specifications.}
	  \scalebox{0.8}{
	\begin{tabular}{lr} \hline	
		Effective photo-sensitive area & 3~mm $\times$ 3~mm \\ 
		Pixel pitch & 50~$\mu$m \\
		Number of pixels &  3600 \\ 
		Number of channels & 1ch\\
		Package & Metal with TE-cooler \\
		Dark count (at 10$^\circ$C) & 200~kHz\\
		Gain & $8 \times10^{5}$ \\ 
		\hline		
	\end{tabular}}
	\label{tab:MPPC}
	\end{center}
	\end{minipage}
\end{table}

Each detector strip is composed of BC404 plastic scintillator, manufactured by Saint-Gobain~\cite{SAINTGOBAIN},
Y11~(350) WLS fiber by Kuraray~\cite{KURARAY}, and S10943-0928(X) MPPCs by Hamamatsu Photonics~\cite{HAMAMATSU}.
Seven 1-mm-diameter fibers are embedded and glued into grooves on each scintillator strip at $10~\mathrm{mm}$ intervals.  
The cross section of a groove
\footnote{The bottom of the groove is a circular shape to make a full contact between the surface of the fibers and the scintillator.}
 is $1.1~\mathrm{mm} \times 1.1~\mathrm{mm}$.
To achieve a readout at both ends without interfering
with a neighboring quadrant, the fibers are not glued in the $20$-$\mathrm{mm}$ region at the end of the strip and are fed to the opposite side of the CFRP plate through holes, as shown in Fig.~\ref{fig:CVPhoto}.
The fibers are bundled at both ends and connected to the MPPCs, which are mounted on the outer supporting structure.
Each scintillator strip is wrapped with a reflector film: an aluminized polyester film (Tetolight) manufacured by OIKE \& Co., Ltd.~\cite{OIKE}.
The thicknesses of the base film and the deposited aluminum are $12~\mu\mathrm{m}$ and $40$-$50~\mathrm{nm}$, respectively.

Table~\ref{tab:MPPC} shows specifications of the MPPC. 
In order to stabilize the gain and quantum efficiency and reduce dark noise,
a Peltier cooler is used to keep and control the temperature of the MPPCs to be $10 \pm\ 0.1\ ^{\circ}\mathrm{C}$. 
For these purposes, we developed a 3-mm-square MPPC, which is integrated with a Peltier device, jointly with Hamamatsu Photonics.
\subsection{Production of the CV strip}
The production of the scintillator strip had five steps: cutting a scintillator plate, cutting grooves, embedding fibers, bundling fibers at each end into a connector, and wrapping the strip with a reflector film.
We first measured the thickness of sixteen raw scintillator plates over the whole area of $1010~\mathrm{mm} \times 570~\mathrm{mm}$
by a laser displacement meter to determine how to cut strips out of the plates.
The mean thickness was $3.10~\mathrm{mm}$ and the standard deviation was $0.19~\mathrm{mm}$.
The measurement accuracy was $\pm 0.03~\mathrm{mm}$.
The region whose thickness was less than $2.75~\mathrm{mm}$ were not used because the light yield under the grooves was too low and it would increase the inefficiency.
After cutting strips out of the plates, we made seven grooves for fibers by a milling machine. 
Next, we embedded fibers into the grooves and glued them by using an automatic applicator of an optical cement (ELJEN EJ-500~\cite{ELJEN}).
We then bundled the fibers at each end into a connector made with aluminum.
The connector ensures optical coupling with the MPPC;
it keeps the distance between the end of fibers and the surface of the MPPC at $0.6~\mathrm{mm}$
\footnote{There is a protective resin film ($0.3\pm0.2~\mathrm{mm}$ in thickness) upon the surface of MPPC; we made the distance as short as possible.}.
The light loss at the coupling was estimated by a simulation to be 10\% in nominal case, 15\% at most when a possible misalignment was considered.
We glued the fibers with the connector with the optical cement 
and polished the end surface of the fiber bundle by a diamond polisher.
Finally, we wrapped each scintillator strip with the reflector film, which was described in the previous section.

\begin{figure}[!h]
\begin{center}
	 \includegraphics[width=10cm,clip]{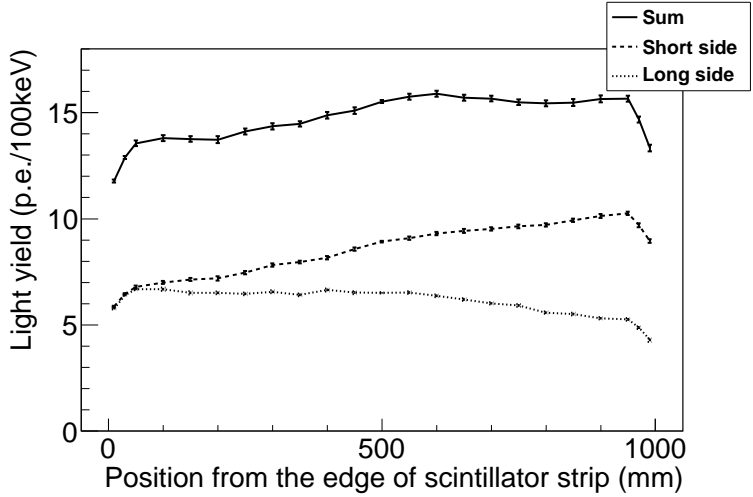}
		\caption{Example of the light yield distribution as a function of the position of the $^{90}$Sr $\beta$ source.
		The horizontal axis indicates the distance of the source from the ``long side'' end of the strip. 
		The configuration of ``short side'' and ``long side'' were mentioned in the caption of Fig.~\ref{fig:CVDesign}.
		The measured light yield was attenuated along the distance from the readout.
		The light yield was low in the regions where the fibers are not embedded in the scintillator.
		}
	\label{fig:CVLight}
	\end{center}
\end{figure}

After the production of the strips, we checked their light yield by using a $^{90}$Sr $\beta$ source.
Figure~\ref{fig:CVLight} shows the light yield of each scintillator strip as a function of the $\beta$ source position.
Here the light yield at both ends and the sum of the two are presented in units of the number of photoelectrons per $100~\mathrm{keV}$ energy deposit.
The energy deposit of the $\beta$ ray was estimated by a simulation.
We obtained the light yield of all the strips to be more than $10~\mathrm{p.e.}/100~\mathrm{keV}$ in the whole region.
For every group of four scintillator strips with the same shape and size in each layer(note that the CV has a 90-degree symmetry), 
the light yields at the center of the strips matched with 2\% in $\sigma$ after correcting for thickness variations.
We succeeded in the production of high-light yield scintillator strips with good quality.

\subsection{MPPC readout}
\begin{figure}[!h]
\begin{center}
	 \includegraphics[width=12cm,clip]{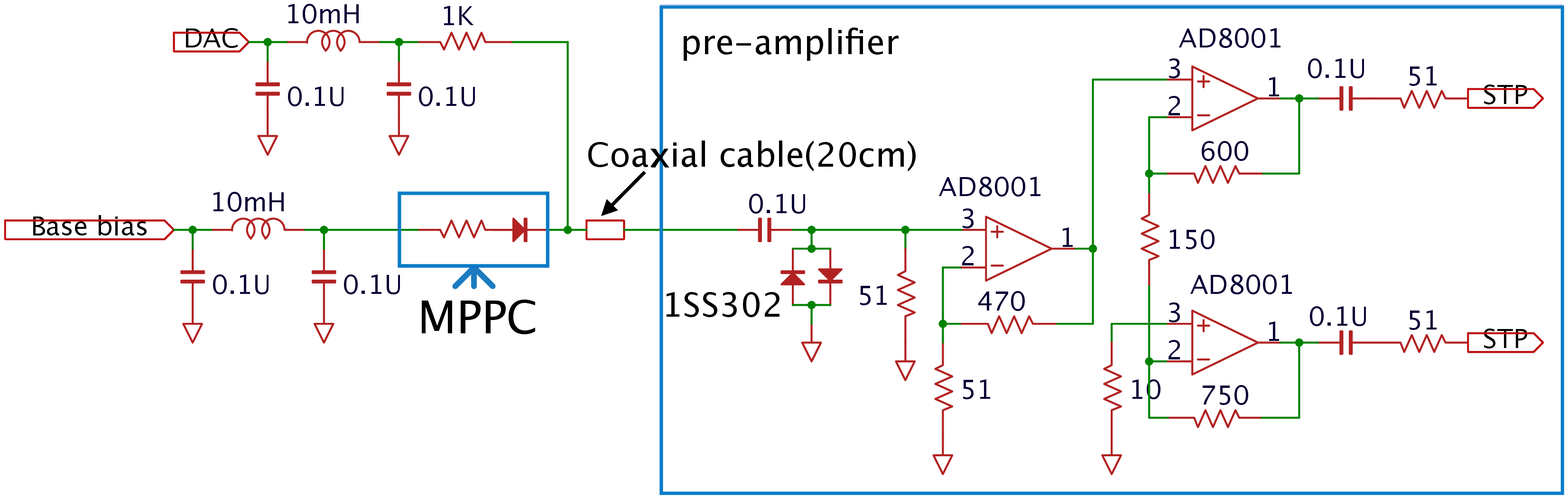}	
	\caption{Schematic of the MPPC bias and readout circuit. 
	The left part indicates the MPPC and the bias voltage connection, and the right part indicates the preamplifier circuit. 
	}
	\label{fig:CVReadout}
	\end{center}
\end{figure}

\begin{figure}[!h]
\begin{center}
	 \includegraphics[width=10cm,clip]{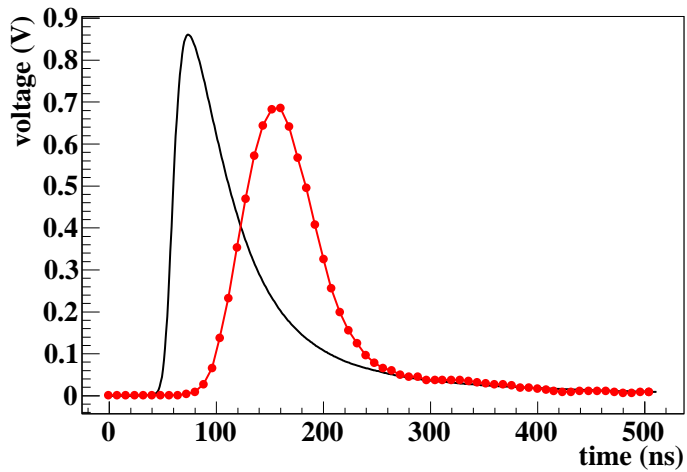}
		\caption{Example of the CV pulse shape of charged particles.
	The smooth line shows an output of the preamplifier.
	The dots indicate data recorded by a 125~MHz ADC module, drawn with a poly-line through the data points.
	Both waveforms were obtained by averaging over 10000 pulses.
		} 
	\label{fig:CVSignal}
	\end{center}
\end{figure}

Figure~\ref{fig:CVReadout} shows the readout circuit for the MPPC.
The anodes of MPPCs are connected to a common power supply, which determines a main bias voltage.
The cathode of each MPPC is connected to a digital-to-analog converter (DAC) that adjusts the bias voltage individually.
A signal from the MPPC cathode is sent to a 50-fold preamplifier through a 20-cm long coaxial cable.
Table~\ref{tab:amp} summarizes the performances of the preamplifier.
The signal, received via an AC coupling, is amplified by a 10-fold non-inverting amplifier and a 5-fold differential amplifier, and is sent to a 125-MHz ADC module~\cite{ADC} through a shielded twisted pair (STP) cable.
The ADC shapes a signal waveform by a 10-pole Bessel filter and records the pulse heights every 8 ns for 64 points (Fig.~\ref{fig:CVSignal}).
In the KOTO experiment, the information of waveform are used to resolve multiple pulses caused by high counting rates.
\begin{table}[!h]
	 \begin{center}
	  \caption{Summary of the performances of the MPPC preamplifier. These values were measured with a 51 $\Omega$ impedance.}.
	\begin{tabular}{lr} \hline
		Output dynamic range & 2.5~V \\ 
		Bandwidth & 200~MHz  \\
		Gain & 51 \\
		Power consumption & 150~mW\\
		Noise & 2.4~mV (RMS) \\ \hline
	\end{tabular}
		\label{tab:amp}
	\end{center}
\end{table}

The combined gain of the MPPC and the preamplifier is monitored by checking a pulse height distribution for minimum ionizing particles (MIPs) passing through the CV layer.
We also monitor the output charge distribution for one photoelectron gain.
By evaluating the MIP peak in units of photoelectron, we monitor the light yield of the CV strip.
It monitors both light outputs from the strip, and the quantum efficiency of the MPPC.

\section{Performance study of a prototype strip}
\begin{figure}[!h]
\begin{center}
	 \includegraphics[width=14cm,clip]{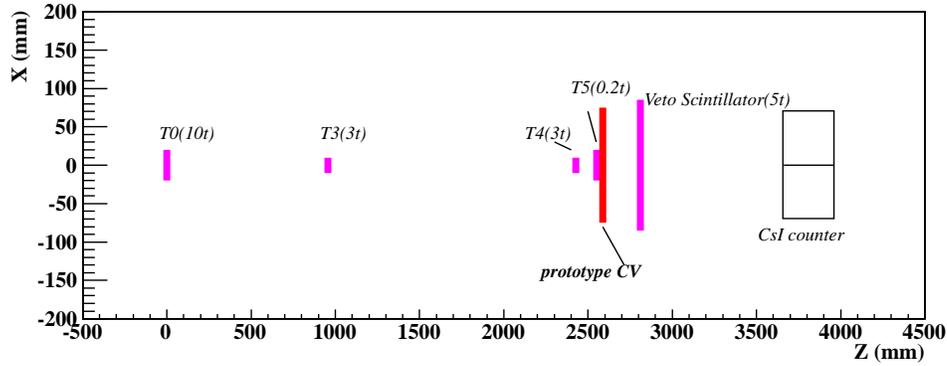}
		\caption{Plan view of the setup for the performance study at ELPH at Tohoku University.
		Four plastic scintillators (T0, T3, T4, and T5) defined the positron beam,
		while another scintillator (Veto scintillator) and an electromagnetic calorimeter (CsI counter) were located behind the prototype strip to identify annihilation events.
		}
	\label{fig:ISetUp}
	\end{center}
\end{figure}

Before the mass production of the CV strips, we evaluated the performance of a prototype strip
by using a $600~\mathrm{MeV}/c$ positron beam at the Research Center for Electron Photon Science (ELPH) at Tohoku University. 
Figure~\ref{fig:ISetUp} shows the experimental setup for the study. 
Four plastic scintillators (T0, T3, T4, and T5
\footnote{
To reduce the effect of the annihilation which occur in the upstream of the CV,
we prepared a thin (0.2 mm thick) plastic scintillator, T5,  just in front of the CV.
We can select charged particles at T5 with a small probability of an annihilation in itself. 
The contamination of the annihilation in T5 to the whole annihilation events were 20$\%$, based on a simulation.}
) were used to define the beam size ($20~\mathrm{mm} \times 20~\mathrm{mm}$) and ensured that the positrons were hitting the prototype strip.
Another plastic scintillator (Veto Scintillator) and an electromagnetic calorimeter (CsI counter) were located downstream of the prototype CV to be tested, 
in order to identify positron annihilation events.
We defined an event with energy deposits in the calorimeter without a hit in the veto scintillator as an annihilation event. 
The calorimeter was a $2 \times 2$ array of CsI crystals with dimensions of 7-cm square in cross section and $30~\mathrm{cm}$ in length.

We used a $15$-$\mathrm{cm}$ long and $7$-$\mathrm{cm}$ wide prototype strip.
In this study, we used a different readout system than the one described in Sect. 2.4, which includes a commercial amplifier module for MPPC and a charge integrating ADC for recording the signal.

\begin{figure}[!h]
\begin{center}
	 \includegraphics[width=10cm,clip]{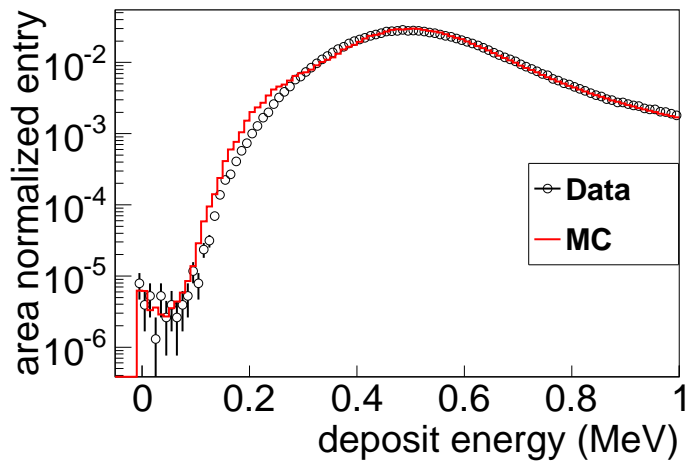}
	 \caption{Energy deposit in the prototype strip. 
		The energy scale of the data was determined so that the peak values in the data and the simulation matched each other.
		The spectrum of the data was normalized to the area over a full energy range.
		The simulated spectrum was also normalized in the same way.
		The energy spectrum of the simulation was smeared with the measured light yield in this measurement (6.8 p.e./100 keV).	
		}
	\label{fig:ISpect}
	\end{center}
\end{figure}

Figure \ref{fig:ISpect} shows the energy deposit distribution of the prototype.
The result from our Monte Carlo (MC) simulation is also presented in the figure.
According to the simulation study, energy deposits by MIPs peak around 0.5~MeV, and
a feature in the region below 0.25~MeV have contributions from both the fluctuation due to the photoelectron statistics
\footnote{The light yield of the prototype strip was less than half of the achieved light yield in the production strips due to a bad optical connection between the WLS fibers and the MPPC.},
and the effect of grooves for WLS fibers, where the effective scintillator thickness is thinner than in other places.
The beam size defined with the plastic scintillators can contribute to the discrepancy 
between the data and the simulation around 0.25 MeV in Fig.~8,
because it changes the contribution of the groove region.
We defined that the prototype strip has a hit if it has more than 100 keV energy deposit.
Table~\ref{tab:SummaryAn} summarizes the results of the inefficiency measurement.
Based on 41 events below the 100 keV threshold, 
the inefficiency of the prototype strip was determined to be $5.4 \times 10^{-5}$, which agreed well with the simulation result.

\begin{figure}[!h]
\begin{center}
	 \includegraphics[width=10cm,clip]{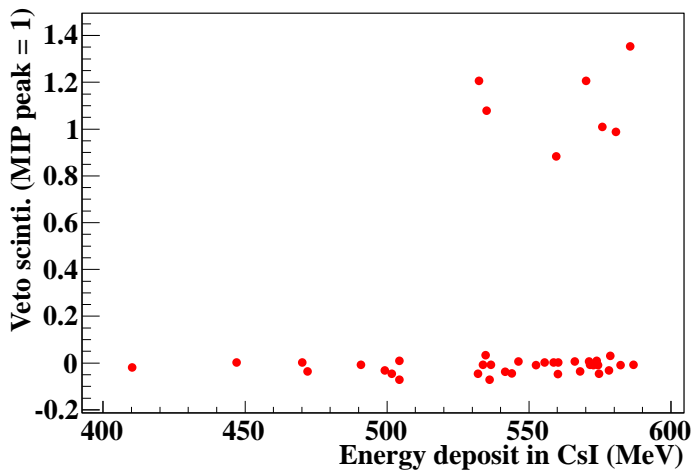}
		\caption{Correlation between the energy deposits in the CsI counter and the veto scintillator.
		The energy deposit in the veto scintillator is normalized by the MIP peak value.
		}
	\label{fig:CsISpect}
	\end{center}
\end{figure}

We investigated the events below 100~keV which were defined as inefficient events.
Figure~\ref{fig:CsISpect} shows the correlation between the energy deposits in the CsI counter and the veto scintillator.
Among 41 events, seven events had energy deposits both in the CsI counter and the veto scintillator, 
and were categorized as the inefficiency for truly penetrating tracks due to the fluctuation of the photoelectron statistics
\footnote{This type of inefficiency is expected to be small given the achieved light yield of the production strips.}.
The other 34 events had energy deposits only in the CsI counter, which means only photons existed behind the prototype strip,
and therefore were categorized as the inefficiency due to positron annihilation in the strip. 
The number of simulated inefficient events was calculated from the information of the particle which hit the prototype CV.
In the simulation, 99\% of the ``photoelectron statistics" were due to the lower deposit energy under the grooves, and has the uncertainty due to the beam size as mentioned before
\footnote{In the simulation, such uncertainty  was 50\%.}. 
On the other hand, the ``annihilation" had no influence from the beam profile.
The resulting inefficiencies due to photoelectron statistics and annihilation are $(0.9 \pm 0.3) \times 10^{-5}$ and $(4.5 \pm 0.8) \times 10^{-5}$, respectively,
and agreed with the MC within the error including the statistical error and the systematic effect from beam size.
We concluded that we had a good understanding of the inefficiency in the positron detection.
\begin{table}[!h]
\begin{center}
	\caption{Summary of the inefficiency measurement of the prototype strip.
	Only statistical errors were considered in the inefficiency values.}
	\begin{tabular}{lcc} \hline
		& Data & Simulation\\  \hline
		N (total events)  &  $7.6 \times 10^5$&$1.04 \times 10^8$\\ 
		N ($<100$~keV)  & 41 & 6318 \\
		Annihilation Inefficiency & $(4.5 \pm 0.8) \times 10^{-5}$ & $(4.0 \pm 0.06) \times 10^{-5}$\\ 
		Photoelectron Statistics Inefficiency & $(0.9 \pm 0.3) \times 10^{-5}$ & $(2.1 \pm 0.05) \times 10^{-5}$\\ 
		Total Inefficiency & $(5.4 \pm 0.8) \times 10^{-5}$ & $(6.1 \pm 0.08) \times 10^{-5}$\\ \hline
	\end{tabular}
	\label{tab:SummaryAn}
\end{center}
\end{table}

\section{Performance study at the beam line for the KOTO experiment}\label{Ptest}
After completing the CV construction, we evaluated the performance of the CV at the neutral beam line for the KOTO experiment: 
the KL beam line in J-PARC Hadron Experimental Facility (HEF).
The measurement was carried out before the full installation of the KOTO detector.
In this measurement, we evaluated the position dependence of the light yield, the timing resolution,
and the inefficiency for charged particles that penetrate the CV layers.
The charged particles were the particles from $K_L \rightarrow e^\pm \pi^\mp \nu$ decay,  $K_L \rightarrow \mu^\pm \pi^\mp \nu$ decay and $K_L \rightarrow \pi^+ \pi^- \pi^0$ decay,
and the mean momentum of these was $1~\mathrm{GeV}/c$.
From the simulation, 59\% of the charged particles were charged pions, 24\% of these were muons, and 17\% of these were electrons and positrons.

\begin{figure}[!htb]
\begin{center}
	 \includegraphics[width=14cm,clip]{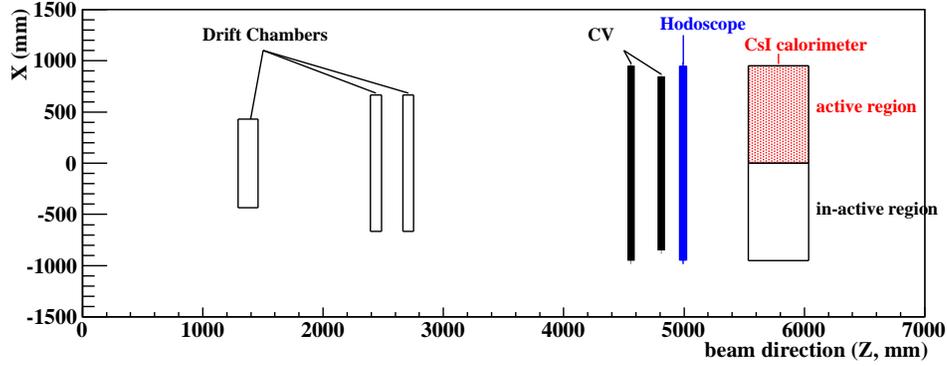}
		\caption{Setup of the performance study at the KL beam line.
		Three sets of drift chambers were used to track charged particles, and a hodoscope placed downstream of the CV layers was used to trigger events.
		}
		\label{fig:PSetUp}
	\end{center}
\end{figure}

\begin{figure}[!htb]
	\begin{center}
	\includegraphics[width=8cm]{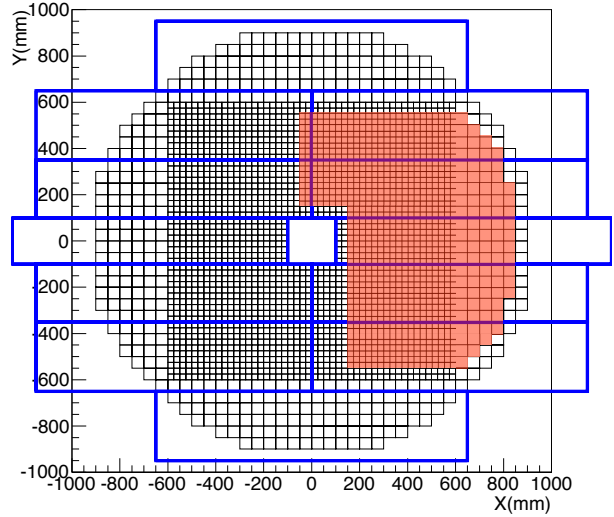}
		\caption{Rear view of the trigger hodoscope and the calorimeter.
		Twelve scintillator plates, represented as rectangular boxes, covered the whole region of the CV and the calorimeter.
		The two sizes of squares in the array show the small and large CsI crystals in the calorimeter.
		In this study, only the region of the calorimeter shown in red was used for the inefficiency measurement.
		}
		\label{fig:HodCsI}
	\end{center}
\end{figure}

Figures~\ref{fig:PSetUp} and \ref{fig:HodCsI} show a plan view of the experimental setup, and a rear view of the hodoscope and the calorimeter, respectively.
The CV layers were located in front of the CsI calorimeter as in the case of the KOTO experiment
\footnote{Distances from the calorimeter surface were slightly different in this study from those in the KOTO experiment.}.
The CsI calorimeter in the KOTO experiment consists of 2240 small and 476 large undoped CsI crystals.
Their cross sections are 25-mm square and 50-mm square, respectively, and the length is 500~mm.
Only half of the area of the calorimeter was available in this measurement, as shown in Fig.~\ref{fig:HodCsI}. 
A trigger hodoscope, which consisted of twelve 10-mm-thick plastic scintillators, was installed between the CV and the CsI calorimeter.
Three sets of drift chambers were placed upstream of the CV to track charged particles.
Each set of the chambers measured X and Y positions of the track separately.
The pointing resolution of the chamber tracking at the CV layers was approximately $2~\mathrm{mm}$.
The beam power of the J-PARC main ring accelerator for delivery to HEF was $6~\mathrm{kW}$ at the time of this study.
All counters in this study were located in air, while in the KOTO experiment most of the detectors, including the CV and the calorimeter,
are operated in vacuum. 

We required two hits in the hodoscope to collect events with two tracks penetrating the CV layers.
In addition, at least one hit was required in each quadrant of the hodoscope which were a diagonal pair: ($-$X, $+$Y) and ($+$X, $-$Y), or ($-$X, $-$Y) and ($+$X, $+$Y) regions.
In the analysis, tracks were reconstructed with the hits in the chambers, 
and then a corresponding hit in the hodoscope was required for each track to ensure that the particle penetrated the CV.
Further information on the track was obtained by the fine-segmented CsI calorimeter.
In the inefficiency measurement, the matching between the track extrapolation to the calorimeter and the hit location on the calorimeter was checked
to eliminate the events in which a charged particle scattered or decayed in flight.
This limited the area for the inefficiency measurement to be about half of the total CV region, corresponding to the active area of the CsI calorimeter.

\subsection{Evaluation of the light yield}
We evaluated the light yield and determined that the CV satisfies our requirement: more than $10~\mathrm{p.e.}/100~\mathrm{keV}$ in the entire area within the 850~mm radius.
We first reconstructed two tracks through the chambers. 
From the measured X and Y positions, the tracks in the X-Z and Y-Z planes were reconstructed independently,
where Z denotes the coordinate along the beam direction.
To resolve the ambiguity of the combination of tracks, 
we required hits in the hodoscope corresponding to the combination. 

\begin{figure}[!h]
\begin{center}
	 \includegraphics[width=10cm,clip]{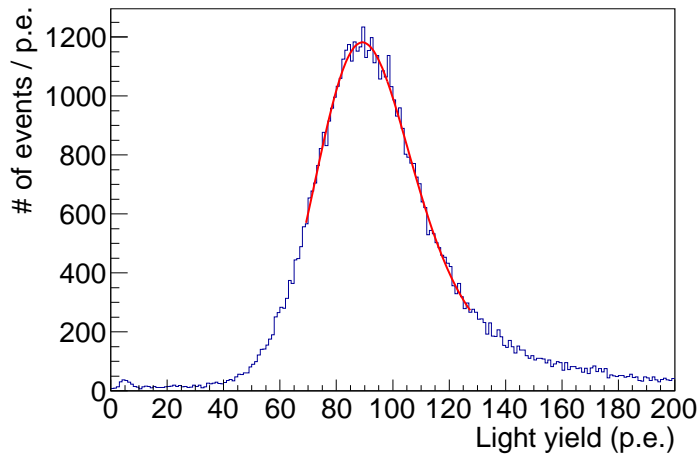}	
		\caption{Example of the light yield distribution of a CV strip.
		The histogram indicates the data, and the curve around the peak shows the fit corresponding to the convoluted function of Landau and Gaussian distributions. 
		The events around $0~\mathrm{p.e.}$ are due to scattering or decay in flight after the chambers.
		}
		\label{fig:LE}
	\end{center}
\end{figure}

\begin{figure}[!h]
\begin{center}
	\includegraphics[width=14cm,clip]{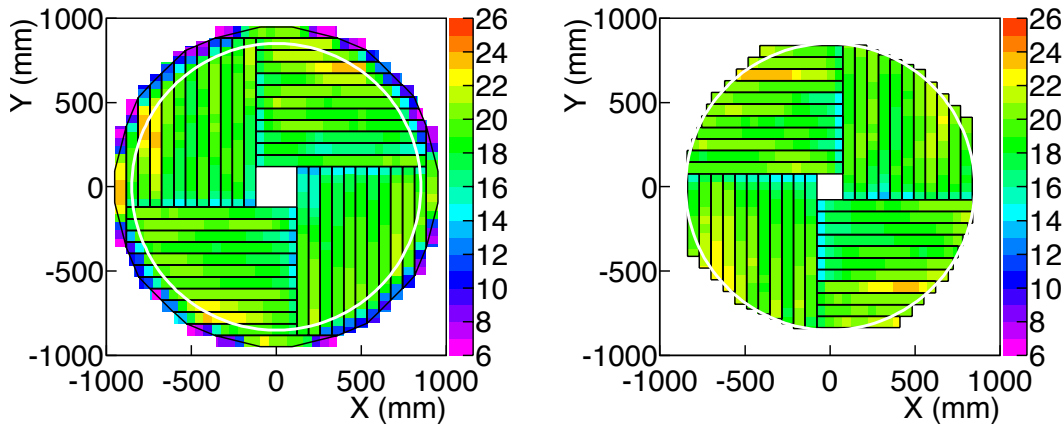}
	\caption{Position dependence of the light yield. 
	The left and right figures correspond to the results of the CV front and rear layers, respectively.
	Color indicates the light yield in units of $\mathrm{p.e.}/100~\mathrm{keV}$ as shown in the reference in the figure.
	The white circle shows a radius of $850~\mathrm{mm}$.
 	Smaller yields were observed at the ends of strips, which were due to smaller light collection in the regions where the fibers were not embedded in the scintillator
	(The outer ends of the strips in the rear layer do not have these regions  of reduced light yield and thus no decreases were observed).
	Because the length of the WLS fibers for the long side at the outer strips is shorter than that at the inner strips due to the WLS fibers routing, the light yield at the outer region tend to be higher than that at the inner region.
	Higher light yield in some outer strips came from fluctuations in the scintillator thickness.
	}
	\label{fig:PosiL}
	\end{center}
\end{figure}

The light yield of each CV strip was obtained as follows:
we integrated a signal waveform, recorded by the $125$-$\mathrm{MHz}$ ADC module, over the range of 64 samples ($512~\mathrm{ns}$) and obtained the output charge.
Next, we converted the output to the equivalent value in photoelectron units.
The output charge for one photoelectron was derived for each MPPC from the charge distribution of dark noise,
which was dominated by the thermo electrons.
We also made a correction for the path length in the strip.
The output was scaled using the incident angle to the value in normal incidence
\footnote{Incident angles of charged particles in this study were less than 20$^\circ$, and thus the correction was a few\% effect.}.
We then summed the outputs from both ends of the strip and defined it as the light yield of the strip.
To evaluate the light yield as a function of the track hit position,
we used the average value in each 5~cm region in the longitudinal direction of the strip.
Figure~\ref{fig:LE} shows an example of the light yield distribution.
We applied a fit to the distribution, which was a convolution of Landau and Gaussian distributions.
We defined the MIP peak as the most probable value in the resultant function.
Corresponding energy deposit to a MIP peak was obtained by the simulation, and the light yield value was normalized by energy deposit in units of $\mathrm{p.e.}/100~\mathrm{keV}$.
Figure \ref{fig:PosiL} shows the position dependence of the light yield.
The light yield exceeded $10~\mathrm{p.e.}/100~\mathrm{keV}$ in the whole region within a $850~\mathrm{mm}$ radius, and the average value over the region was $18.6~\mathrm{p.e.}/100~\mathrm{keV}$.

\subsection{Evaluation of the timing resolution}
\begin{figure}[!h]
\begin{center}
	 \includegraphics[width=14cm,clip]{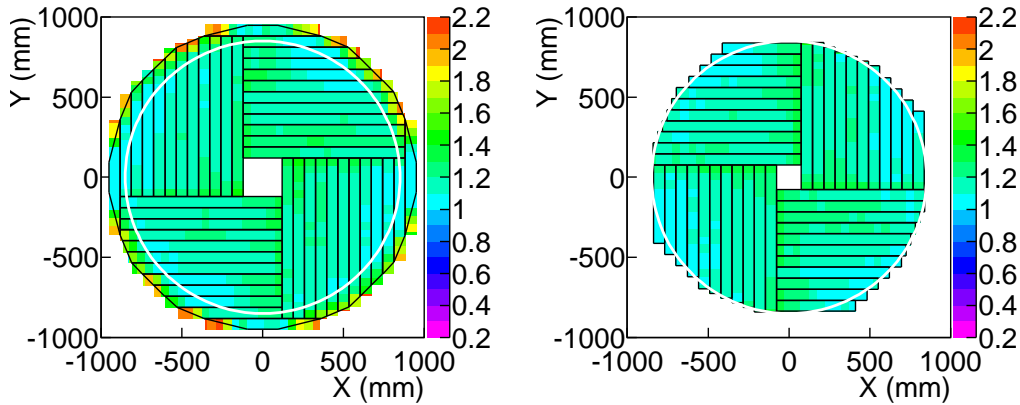}
		\caption{Position dependence of the timing resolution ($\sigma_{\rm Mean}$).
		The left and right figures correspond to the results of the CV front and rear layers, respectively.
		Color indicates the timing resolution in units of ns as shown in the reference in the figure.
		}	
		\label{fig:CVTimePosi}
	\end{center}
\end{figure}

We evaluated the timing resolution and determined that the CV satisfies our requirement: better than $3~\mathrm{ns}$ for the entire area within the $850$-$\mathrm{mm}$ radius.
Similar to the light yield study, we obtained the results for tracks in each 5-cm region in the longitudinal direction of the strip.
The hit timing at each end was calculated by using a constant fraction method;
from a waveform recorded in the $125$-$\mathrm{MHz}$ ADC module, gaps between neighboring samples were  interpolated by a linear function,
and the time where the signal height reached the half of the maximum height in the leading edge was determined.
In this study, we used the difference in the hit timings at both ends in order to eliminate the fluctuation due to a timing resolution of the trigger hodoscope,
while the hit timing in the KOTO experiment is defined as a mean of hit timings at both ends.
In an ideal case, the mean time and the time difference satisfy the equation:
\begin{eqnarray} 
 \sigma_{\rm Mean} =  \frac{ \sigma_{\rm Difference}}{2} = \frac{\sqrt{\sigma^2_{\rm Short} + \sigma^2_{\rm Long}}}{2} \; ,
\end{eqnarray}
where $\sigma_{\rm Mean}$, $\sigma_{\rm Difference}$, $\sigma_{\rm Short}$, and $\sigma_{\rm Long}$ show the root-mean-square of the distributions of
the mean time, the time difference, and the hit times at both ends (short and long sides), respectively.
We calculated the expected timing resolution in the KOTO experiment ($\sigma_{\rm Mean}$) by dividing $\sigma_{\rm Difference}$ by two
\footnote{
The time difference depends on the hit position since the propagation times to both ends change, the resolution here included this contribution.
The contribution to $\sigma_{\rm Mean}$ in the case of a $5$-$\mathrm{cm}$ region was estimated to be $0.1~\mathrm{ns}$ and was disregarded in this study.
}.
Figure~\ref{fig:CVTimePosi} shows a position dependence of the timing resolution.
Comparing it with Fig.~\ref{fig:PosiL}, the timing resolution is larger in the region where the light yield was small.
This is due to the fact that a smaller light yield caused a larger time jitter of the signal pulse.

The timing resolution is smaller than $3~\mathrm{ns}$ in the entire region within the 850~mm radius, 
and the average timing resolution over the region is $1.2~\mathrm{ns}$.

\subsection{Evaluation of the inefficiency}
The analysis of the inefficiency consisted of four steps:
1) event topology selection,
2) candidate track selection,
3) CV hit decision,
and 4) correction.

\paragraph{Event topology selection}
First, tracks were found using the drift chambers.
Then a corresponding hit in the trigger hodoscope was required for each of the two candidates,
as described in Sect.~\ref{Ptest}. 
Events were further required to have a topology with one track going into the CsI active region (``near-track'') and the other into the opposite side (``far-track'').
To ensure the correct combination of the tracks in the X-Z and Y-Z planes, the far-track needs to have corresponding hits in the CV strips.
Only the near-tracks were used to measure the inefficiency.

\begin{figure}[!h]
 \begin{center}
	 \includegraphics[width=12cm,clip]{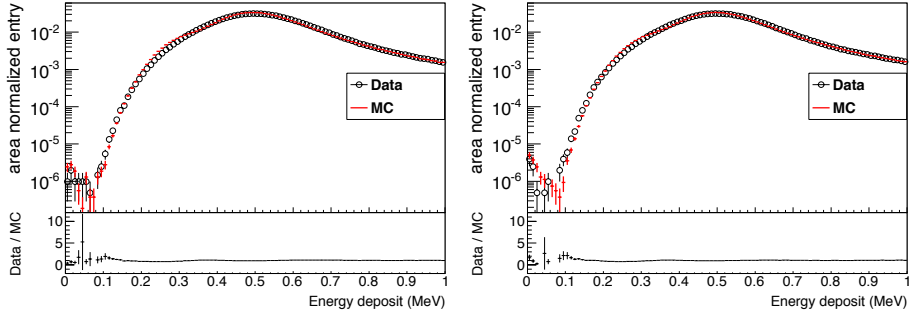}
	\caption{Energy deposit distribution of the CV layers after applying all the cuts.
		The left and right figures correspond to the results of the front and rear layers, respectively.
		The dots show the data and the histogram shows simulation results.
		These spectra were normalized as the same way as in Fig. \ref{fig:ISpect}. 
		The small discrepancy in the region around 300 keV were considered to be due to the reproducibility of the response at the region where the WLS fibers were embedded.
		In addition, the discrepancy at the region lower than 150 keV were considered to be due to the reproducibility at the boundary of the scintillator.
		The light collection at the boundary is smaller.
	}
	\label{fig:CVEne}
	\end{center}
\end{figure}

\paragraph{Candidate track selection}
The second step selected candidate ``near-tracks'' which ensured penetration into the CV. 
Our major concern was charged particles which decay in flight or scatter in the materials between the chamber and the CV,
so that they actually miss the strip which was pointed to by the reconstructed track. 
These ``non-penetrating" tracks could fake inefficient events, and were removed by applying the following cuts. 
First, a CsI hit was requested at the position which matched the track (track-CsI matching cut).
The distance $\Delta R$, between the position of the CsI hit and the chamber tracks, had to be $\Delta  R < $30~mm
\footnote{
This cut value, 30~mm, was chosen considering the resolution of $\Delta  R$ ($\sim$16~mm) which was mainly due to the position resolution of the calorimeter to the charged particles.}.
The CsI hit position was defined as the center of energy deposits in the crystals that formed a cluster around the particle's incident position.
Since the active area of CsI was smaller than that of CV, the track-CsI matching cut 
restricted, in effect, the track position to be within a 720~mm in radius (outer fiducial cut).
Second, tracks were rejected when the expected position at the CV was closer than 200~mm from the beam center (inner fiducial cut).
This cut was useful to remove those particles that decayed or scattered into the beam hole
\footnote{
Intuitively, it could be understood as the sum of $\Delta  R$ (30~mm) and half of the size of the maximum length of the beam hole (diagonal direction$\sim$170~mm).
}.
These cuts described above were found to remove non-penetrating tracks almost entirely.
\paragraph{CV hit decision}
The third step selected hits in the CV.
Given a near-track which passed the cuts in the second step, 
we examined those CV strips that had an overlap with a virtual square set on the CV plane 
whose center was placed at the track's expected position.
The size of the square was chosen to be 14~cm $\times$14~cm considering the width of the CV strips, and 
the number of the examined strips was 2 to 5, depending on the track position.
We then took the maximum energy deposit among these strips as the CV energy deposit for the track under consideration.
To check the validity of the each selection, we carried out the MC simulation which takes into account the response of the detectors (the chambers, CV, hodoscope, CsI, in-beam material).
We simulated the three main decay modes: $K_L \rightarrow e^\pm \pi^\mp \nu$,  
	$K_L \rightarrow \mu^\pm \pi^\mp \nu$, and $K_L \rightarrow \pi^+ \pi^- \pi^0$.
Figure~\ref{fig:CVEne} shows the energy distribution of data and simulation determined by this method. 
The agreement between data and simulation was good.
The simulation was also used for correcting the inefficiency as mentioned later.
The CV is defined to have a hit if it has an energy deposit more than 100 keV,
	and CV inefficiency was defined as the fraction of tracks which had no hits.
We denote this inefficiency by $I_{\rm raw}$.
In the data,
the inefficiencies were $I_{\rm raw(f)}^{\rm data}=(1.13 \pm 0.24)\times 10^{-5}$ and $I_{\rm raw(r)}^{\rm data}=(1.43 \pm 0.26) \times 10^{-5}$, respectively.
The errors indicate statistical uncertainties.
In the simulation,
the estimated inefficiencies were $I_{\rm raw(f)}^{\rm MC}= (1.36 \pm 0.16) \times 10^{-5}$ 
	and $I_{\rm raw(r)}^{\rm MC}=(1.64 \pm 0.18) \times 10^{-5}$, respectively.
The uncertainty is the statistical error.
The inefficiency of data and MC agreed well.
	
\begin{figure}[!h]
\begin{center}
	 \includegraphics[width=12cm,clip]{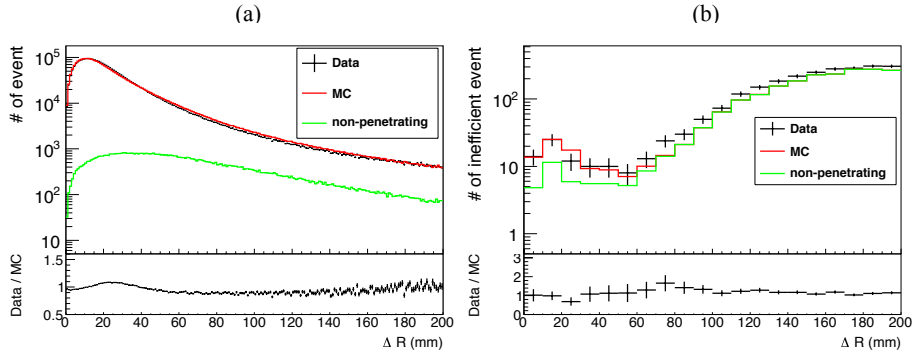}
		\caption{
		(a) Distribution of  ``near" tracks as a function of $\Delta R$. 
		(b) Distribution as a function of $\Delta R$ of the tracks that had energy deposits of less than 100 keV at the front layer or rear layer.
		The histogram labeled ``non-penetrating'' shows the events in which an uncharged particle hit the corresponding CV strip,
		according to the true information obtained from the simulation.
		The histogram for the simulation was normalized to 
		the number of measured "near" tracks for (a), and inefficient events for (b), in each plot.}
		\label{fig:IneffDist}
	\end{center}
\end{figure}

\paragraph{Correction}
The fourth step is a correction to the $I_{\rm raw}$ obtained above.
There are two main factors which should be taken into account: one related to non-penetrating tracks and the other related to accidental hits.
	The former would cause the measured inefficiency ($I_{\rm raw}$) to be larger than the true value, 
	while the latter would cause it to be smaller. We consider them in turn below. 
As described above, non-penetrating tacks were removed almost entirely by the track-CsI matching cut.
Some non-penetrating tracks, however, survived the cut; for example, 
	the events which underwent the $K_L \rightarrow \pi^+ \pi^- \pi^0$ decay upstream of the CV could 
        fake ``correct" CsI hits with the photon(s) from $\pi^{0}$.
	Another example is tracks that scattered producing Bremsstrahlung photon(s).
	We relied on MC simulation to estimate the contribution of those residual non-penetrating tracks, which 
	made the apparent CV inefficiency larger.
	We will correct the raw inefficiency later (Eq.\ref{eq:correction}), introducing a variable $C_{\rm n.p}$,
	which is the ratio of non-penetrating tracks to all the tracks in the final selection.
	Figure~\ref{fig:IneffDist} (a) shows
	a distribution of $\Delta R$,
        for the tracks that passed all the cuts in the three steps described above, except for $\Delta R$.
        The data and simulation agree within 18\%.
	We note that in the simulation, non-penetrating tracks could be identified unambiguously (see the histogram in green).
	Figure~\ref{fig:IneffDist} (b) shows a $\Delta R$ distribution for the inefficient tracks with the CV energy deposits of less than 100~keV. 
	An important observation in this figure is that some non-penetrating tracks (shown in green) satisfied the $\Delta R <30$~mm cut.
	Among the 73 (88) inefficient tracks in the front (rear) layer with $\Delta R < 30$~mm, 27 (42) tracks were identified as being from non-penetrating tracks.
	$C_{\rm n.p}$ is $0.37^{+0.19}_{-0.21}$ for the front layer, and  $0.48^{+0.18}_ {-0.19}$ for the rear layer.
	The errors came mainly from three sources : the first one from statistical uncertainty of MC ($\pm0.08$ or $\pm0.09$), 
	the second from the cutout size in the MC which could affect the ratio through the number of inefficient events for penetrating tracks ($^{+0.09}_{-0.07}$ or $^{+0.10}_{-0.07}$)\footnote{
	Machining error of the cutouts was taken into account, when $C_{\rm n.p}$ was calculated in the MC.},
	and the last one from possible modeling errors on penetrating and/or non-penetrating tracks by MC, which was evaluated with separate control samples ($\pm0.15$ or $\pm0.16$)
	\footnote{
	We prepared two kinds of control samples, penetrating enriched and non-penetrating enriched samples, by applying additional cuts for both data and MC.
	Any modeling errors should appear as discrepancies between MC and data  more clearly for those samples. 
	We found they agreed well each other within statistics, and put possible deviations as systematic uncertainties.  
	}.	
 	
Next, we considered the effects caused by accidental hits.  
According to the data, the mean multiplicity of the CV hits among the strips defined by the 14~cm $\times$ 14~cm squares
	was 1.044 for the front layer, and 1.052 for the rear layer.
The probability of the extra activity, $\sim 0.05$, is a rough measure of correction factor for accidentals.  
We elaborated this estimate with the help of MC studies: {\it i.e.}
	we overlaid MC events with real accidental data taken with a random trigger whose rate is proportional to the instantaneous rate, 
        and studied the nature of accidental hits and the rate. 
In addition to the real accidental hits,
stemming from uncorrelated particles in the beam, we found that there are correlated activities caused by, for example, a delta ray and/or a backsplash from the CsI calorimeter.
The probability of the accidental CV hits ($C_{\rm acc}$) that contribute to the efficiency calculation was found to be 
	slightly smaller than the mean hit multiplicity:
        the values are 
	$C_{\rm acc(f)}=$0.031 $\pm$ 0.006 for the front layer and $C_{\rm acc(r)}=$0.043 $\pm$ 0.006 for the rear layer.
The errors came mainly from systematic uncertainties, 
which were evaluated from discrepancies in the CV hit distributions between the data and MC with accidental overlay.

\paragraph{Results of inefficiency measurements}
	The raw inefficiency was corrected by the equation
	\begin{eqnarray} 
		{I}_{\rm corr} =  {I}_{\rm raw}\times(1-C_{\rm n.p})\times(\frac{1}{1-C_{\rm acc}}). 	
 		\label{eq:correction}
	\end{eqnarray}
The results were
	$I_{\rm corr(f)} =(0.74 \pm 0.15 ^{+0.25}_{-0.22})\times 10^{-5}$ and 
        $I_{\rm corr(r)} =(0.77 \pm 0.14 ^{+0.28}_{-0.27})\times 10^{-5}$, 
        for the front and rear layers, respectively, 
        where the first errors are statistical and the second errors are systematic.
	The source of the systematic errors is dominated by those from $C_{\rm n.p}$ and $C_{\rm acc}$; 
        other sources, such as the stability of energy deposit, deviation of each cut were small and were thus neglected.

\begin{figure}[!h]
 \centering
	 \includegraphics[width=12cm,clip]{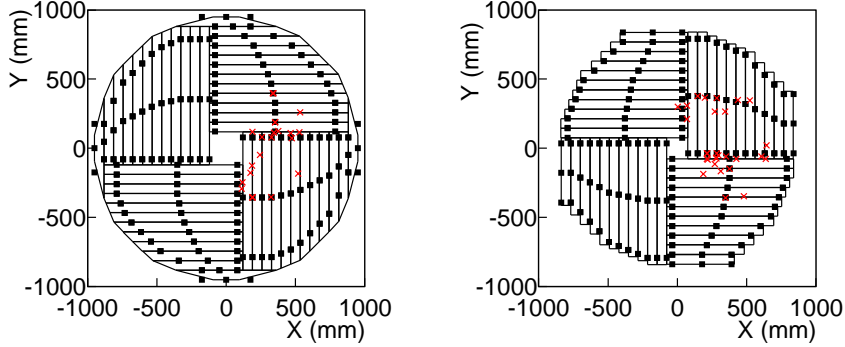}
	\begin{center}
		\caption{Hit positions of the inefficiency tracks in the CV layers.
		The left and right figures show the front and rear layers, respectively.
		The crosses show the positions calculated by the chamber tracking.
		The squares on the edges of the strips indicate the position of the cutouts used to hold the scintillator strips in place.
		The size of the cutouts are magnified 50 times in the plot.
		}
		\label{fig:IneffPosi}
	\end{center}
\end{figure}

        We investigated the hit position of the inefficient tracks on the CV.
	As shown in Fig.~\ref{fig:IneffPosi},
	the majority of the inefficient points were found to be located near the cutouts or on the edges of the scintillator strips, 
        and not in the central region.
       This may be regarded as evidence that the CV was built as designed.
        Finally, we set the upper limit of the inefficiency from $I_{\rm corr}$. 
        Assuming Poisson statistics,  
        the upper limit was found to be 1.5 $\times \ 10^{-5}$(90\%C.L.) and 1.5 $\times \ 10^{-5}$(90\% C.L.)
        for the front and rear layers, respectively.
	We conclude that CV achieved high efficiency($<1.5 \times 10^{-5}$ of inefficiency) and satisfied the requirements for the KOTO experiment.

\section{Conclusion}
We developed a low-mass and high-efficiency charged particle detector for the KOTO experiment at J-PARC.
Two layers of 3-mm-thick plastic scintillators with embedded wavelength-shifting fibers, each of which was read by MPPCs, were manufactured.
The MPPC, which has 3 mm $\times$ 3~mm sensitive area and is equipped with a thermoelectric cooler, was newly developed for this purpose.
After completing the construction, performance studies were conducted at the neutral Kaon beam line for the KOTO experiment in the J-PARC Hadron Experimental Facility.
We measured the light yield, the timing resolution, and the inefficiency against penetrating charged particles.
The resultant light yield per 100~keV energy deposit was 18.6~p.e. on average, 
and the timing resolution of the scintillator strips was measured to be 1.2~ns.
In particular, the detection inefficiency against the particles which penetrated the CV, per layer, was less than $1.5 \times 10^{-5}$ over the area within a radius of 850 mm.
\section*{Acknowledgment}
We are grateful to the staff members of the KEK for their cooperation in the construction of the CV and the measurement of its performance. 
We also thank the staff members of the ELPH accelerator, J-PARC accelerator, Hadron Beam groups and the KEK Computing Research Center
for their supports in taking and analyzing the physics data.
Part of this work was supported by JSPS/MEXT KAKENHI Grant Numbers 23224007 and 8071006.
Some of the authors were supported by Grant-in-Aid for JSPS Fellows.

\end{document}